\documentclass[
    ,final            
  ]
  {aipproc}

\layoutstyle{6x9}

\begin{document}

\title{Precision Electroweak Physics\footnote{Presented at the X Mexican 
Workshop on Particles and Fields, Morelia, Mich., November 6--12, 2005.}}


\author{Jens Erler}{
address={Instituto de F\'\i sica, Universidad Nacional Aut\'onoma de M\'exico,
04510 M\'exico D.F., M\'exico}
}

\begin{abstract}
The status in electroweak precision physics is reviewed. I present a brief 
summary of the latest data, global fit results, a few implications for new 
physics, and an outlook.
\end{abstract}

\maketitle


\section{Observables}
\subsubsection{$Z$ pole}
The $Z$ factories, LEP and SLC, have performed benchmark precision measurements
for the electroweak Standard Model (SM)~\cite{LEPSLD:2005em}. LEP scanned 
the $Z$ lineshape yielding the $Z$ boson mass, $M_Z$, with $2\times 10^{-5}$ 
relative precision, as well as its total width, $\Gamma_Z$, and hadronic peak 
cross section, $\sigma_{\rm had}^0 \equiv 12\pi \Gamma(e^+e^-)\Gamma({\rm had})
/M^2_Z\,\Gamma^2_Z$,
both to better than one per mille accuracy. $\Gamma(\bar{f}f)$ is the $Z$ 
partial decay width into fermion $f$ and $\Gamma({\rm had})$ is the hadronic 
$Z$ decay width. Results on the three leptonic ($\ell = e, \mu,\tau$) branching
ratios, $R_\ell\equiv\Gamma({\rm had})/\Gamma(\ell^+\ell^-)$, are also at 
the per mille level. $\Gamma_Z$, $\sigma_{\rm had}^0$, and the $R_\ell$ are 
unique in their sensitivity to the strong coupling constant, $\alpha_s$, which
can be extracted with very small theoretical uncertainty. The SLC was able 
to compensate its lower luminosity by its electron beam polarization.
The left-right polarization asymmetry, $A_{LR}$, for hadronic final states 
provides the currently most precise value of the weak mixing angle, 
\begin{equation}
  \sin^2\theta_W = {g'^2\over g^2 + g'^2},
\end{equation}
where $g$ and $g'$ are the $SU(2)_L$ and $U(1)_Y$ gauge couplings, 
respectively. Very high precision could also be achieved in the heavy flavor 
sector consisting of branching ratios and various asymmetries for $b$ and $c$ 
quarks. More specifically, the forward-backward (FB) asymmetry into $b$ quarks,
$A_{FB}^b$, amounts to the most precise measurement of $\sin^2\theta_W$ at LEP,
while the few per mille measurement of 
$R_b = \Gamma(\bar{b}b)/\Gamma({\rm had})$ yields independent information on 
the top quark mass, $m_t$, and constraints on new physics affecting the third 
generation in a non-universal way. The heavy flavor results have been finalized
very recently. Analogous results are also available for $s$ quarks, albeit with
larger uncertainties. Other $Z$ pole observables include the three leptonic FB
asymmetries, $A_{FB}^\ell$, the final state $\tau$ polarization and its FB
asymmetry, and charge asymmetry measurements. 

\begin{table}[t]
\begin{tabular}{llccr}
\hline
  & \tablehead{1}{r}{b}{observable}
  & \tablehead{1}{r}{b}{experimental value}
  & \tablehead{1}{r}{b}{SM prediction}
  & \tablehead{1}{r}{b}{pull}   \\
\hline
&$M_Z$          [GeV] & $91.1876\pm 0.0021$  & $91.1874\pm 0.0021$  & $ 0.1$ \\
&$\Gamma_Z$     [GeV] & $ 2.4952\pm 0.0023$  & $ 2.4968\pm 0.0011$  & $-0.7$ \\
&$\sigma_{\rm had}^0$ [nb] & $41.541\pm 0.037$ & $41.467 \pm 0.009$ & $ 2.0$ \\
&$R_e$                & $20.804 \pm 0.050$   & $20.756 \pm 0.011$   & $ 1.0$ \\
&$R_\mu$              & $20.785 \pm 0.033$   & $20.756 \pm 0.011$   & $ 0.9$ \\
&$R_\tau$             & $20.764 \pm 0.045$   & $20.801 \pm 0.011$   & $-0.8$ \\
&$R_b$                & $0.21629\pm 0.00066$ & $0.21578\pm 0.00010$ & $ 0.8$ \\
&$R_c$                & $0.1721 \pm 0.0030$  & $0.17230\pm 0.00004$ & $-0.1$ \\
&$A_{FB}^e$           & $0.0145 \pm 0.0025$  & $0.01622\pm 0.00025$ & $-0.7$ \\
&$A_{FB}^\mu$         & $0.0169 \pm 0.0013$  &                      & $ 0.5$ \\
&$A_{FB}^\tau$        & $0.0188 \pm 0.0017$  &                      & $ 1.5$ \\
&$A_{FB}^b$           & $0.0992 \pm 0.0016$  & $0.1031 \pm 0.0008$  & $-2.4$ \\
&$A_{FB}^c$           & $0.0707 \pm 0.0035$  & $0.0737 \pm 0.0006$  & $-0.8$ \\
&$A_{FB}^s$           & $0.0976 \pm 0.0114$  & $0.1032 \pm 0.0008$  & $-0.5$ \\
&$\bar{s}_\ell^2$     & $0.2324 \pm 0.0012$  & $0.23152\pm 0.00014$ & $ 0.7$ \\
&                     & $0.2238 \pm 0.0050$  &                      & $-1.5$ \\
&$A_e$                & $0.15138\pm 0.00216$ & $0.1471 \pm 0.0011$  & $ 2.0$ \\
&                     & $0.1544 \pm 0.0060$  &                      & $ 1.2$ \\
&                     & $0.1498 \pm 0.0049$  &                      & $0.6$  \\
&$A_\mu$              & $0.142  \pm 0.015$   &                      & $-0.3$ \\
&$A_\tau$             & $0.136  \pm 0.015$   &                      & $-0.7$ \\
&                     & $0.1439 \pm 0.0043$  &                      & $-0.7$ \\
&$A_b$                & $0.923  \pm 0.020$   & $0.9347 \pm 0.0001$  & $-0.6$ \\
&$A_c$                & $0.670  \pm 0.027$   & $0.6678 \pm 0.0005$  & $ 0.1$ \\
&$A_s$                & $0.895  \pm 0.091$   & $0.9356 \pm 0.0001$  & $-0.4$ \\
\hline
\end{tabular}
\caption{$Z$ pole observables compared with the SM best fit predictions. 
$\bar{s}_\ell^2$ is an effective mixing angle which absorbs all radiative 
corrections and corresponds most closely to what enters the $Z$ pole 
asymmetries. The first is extracted from the hadronic charge asymmetry, 
a weighted average over light-quark FB asymmetries. The second is from 
the final state electron FB asymmetry, $A_{FB}$, from CDF~\cite{Acosta:2004wq}.
The three values of $A_e$ are (i) from $A_{LR}$ for hadronic final 
states~\cite{Abe:2000dq}; (ii) from $A_{LR}$ for leptonic final states and from
polarized Bhabba scattering~\cite{Abe:2000hk}; and (iii) from the angular 
distribution of the $\tau$ polarization (LEP)~\cite{LEPSLD:2005em}. The two 
$A_\tau$ values are from SLD and the total $\tau$ polarization, respectively. 
The uncertainties in the SM predictions are induced by the errors in the SM 
parameters, and their correlations have been accounted for.}
\label{tab:zpole}
\end{table}

These $Z$ pole measurements are summarized in Table~\ref{tab:zpole}. Some 
results are quoted in terms of asymmetry parameters, 
\begin{equation}
   A_f \equiv \frac{2 v_f a_f}{v_f^2 + a_f^2}, 
\end{equation}
where at tree level the $Zf\bar{f}$ vector (axial-vector) coupling is given by 
$v_f = T_3^f - 2 Q^f \sin^2\theta_W$ ($a_f = T_3^f$), and where $Q^f$ and $T^f$
denote, respectively, the electric charge and third component of weak isospin
of fermion $f$. The FB asymmetries can also be expressed in terms of these,
$A_{FB}^f = 3/4\, A_e A_f$. 

\begin{figure}[t]
  \includegraphics[height=.47\textheight]{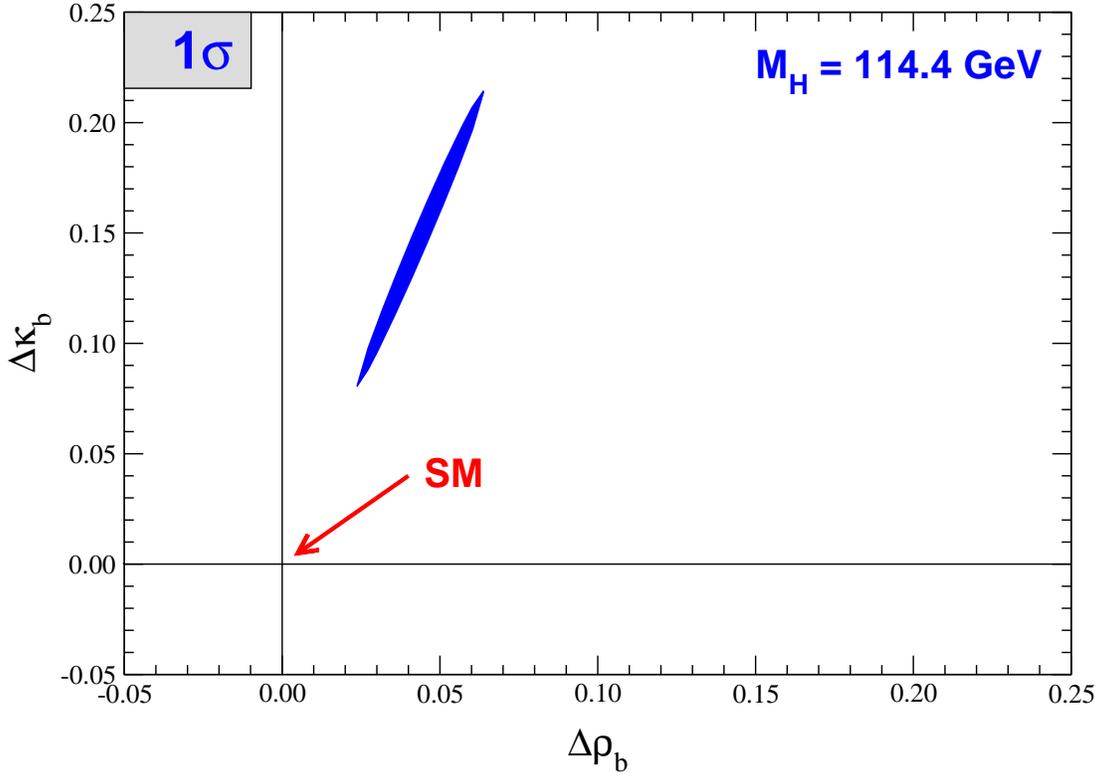}
  \caption{New Physics contributions to form factors for the $Zb\bar{b}$ vector
           and axial-vector couplings.}
\label{fig:bformfactors}
\end{figure}

The pull of an observable gives its deviation from the SM and is defined as,
\begin{equation}
   {\rm pull}(O_i) = {O_i^{\rm exp.}-O_i^{\rm SM}\over\Delta O_i^{\rm total}},
\end{equation}
where $O_i^{\rm exp.}$ is the experimental central value of observable, $O_i$, 
$O_i^{\rm SM}$ is the central value of its SM prediction, and 
$\Delta O_i^{\rm total}$ is the sum in quadrature of the contributing 
statistical, systematical, and theoretical uncertainties, but excludes 
the parametric uncertainty in the SM prediction. As can be seen from 
Table~\ref{tab:zpole}, there are only three $Z$ pole observables which deviate
by two standard deviations ($\sigma$) or more, but interestingly these are all
among the most precise. In particular, $A_{LR}$ and $A_{FB}^b$ 
provide valuable information on the mass of the Higgs boson, $M_H$, and deviate
by 3.1~$\sigma$ from each other. Since $A_{FB}^b$ involves $b$ quarks, and 
the third fermion generation is often suspected to be affected differently by 
physics beyond the SM, one can interpret it alternatively as a measurement of 
$Zb\bar{b}$ couplings. Defining,
\begin{equation}
\begin{array}{l}
   v_b = (1 + \Delta\hat\rho_b + \Delta\rho_b) [T_3^f - 2 Q^f 
   (1 + \Delta\hat\kappa_b + \Delta\kappa_b)\sin^2\theta_W], \vspace{4pt} \\
   a_b = (1 + \Delta\hat\rho_b + \Delta\rho_b) T_3^f,
\end{array}
\end{equation}
one can fit to $\Delta\rho_b$ and $\Delta\kappa_b$, which are due to new 
physics only when all SM contributions are subsumed in $\Delta\hat\rho_b$ and 
$\Delta\hat\kappa_b$. $R_b$ and $A_b$ provide additional constraints. 
The result is shown in Fig.~\ref{fig:bformfactors}, from where it becomes clear
that a correction of 10--20\% to $\Delta\kappa_b$ would be necessary to account
for the data.  This would be a very large radiative correction, given that 
the quadratically enhanced top quark contribution in the SM is less than 1\%. 
It is thus very unlikely that the deviation in $A_{FB}^b$ is due to a loop
effect, but it is conceivably of tree-level type affecting preferentially 
the third generation. Examples include the decay of a scalar neutrino 
resonance~\cite{Erler:1996ww}, mixing of the $b$ quark with heavy 
exotics~\cite{Choudhury:2001hs}, and a heavy $Z'$ with family non-universal 
couplings~\cite{Erler:1999nx}. It is difficult, however, to simultaneously 
account for $R_b$, which has been measured on the $Z$ peak and 
off-peak~\cite{LEPSLD:2005em} at LEP~1. In this context it is interesting that
an average of $R_b$ measurements at LEP~2 at energies between 133 and 207~GeV
is 2.1~$\sigma$ below the SM prediction, and $A_{FB}^b$ is 1.6~$\sigma$ 
low~\cite{LEP:2005di}.

The measurement of $\sigma_{\rm had}^0$ is 2~$\sigma$ higher than the SM 
prediction. As a consequence, when one fits to the number, $N_\nu$, of active
neutrinos\footnote{By definition, an active neutrino is one that couples to 
the $Z$ boson like a standard neutrino.} one obtains a 2~$\sigma$ deficit, 
$N_\nu = 2.986\pm 0.007$, compared to the SM prediction, $N_\nu = 3$. 
Amusingly, LEP~2~\cite{LEP:2005di} also sees a 1.7~$\sigma$ excess in 
the averaged hadronic cross section.

\subsubsection{Other data}
\begin{figure}[t]
  \includegraphics[height=.47\textheight]{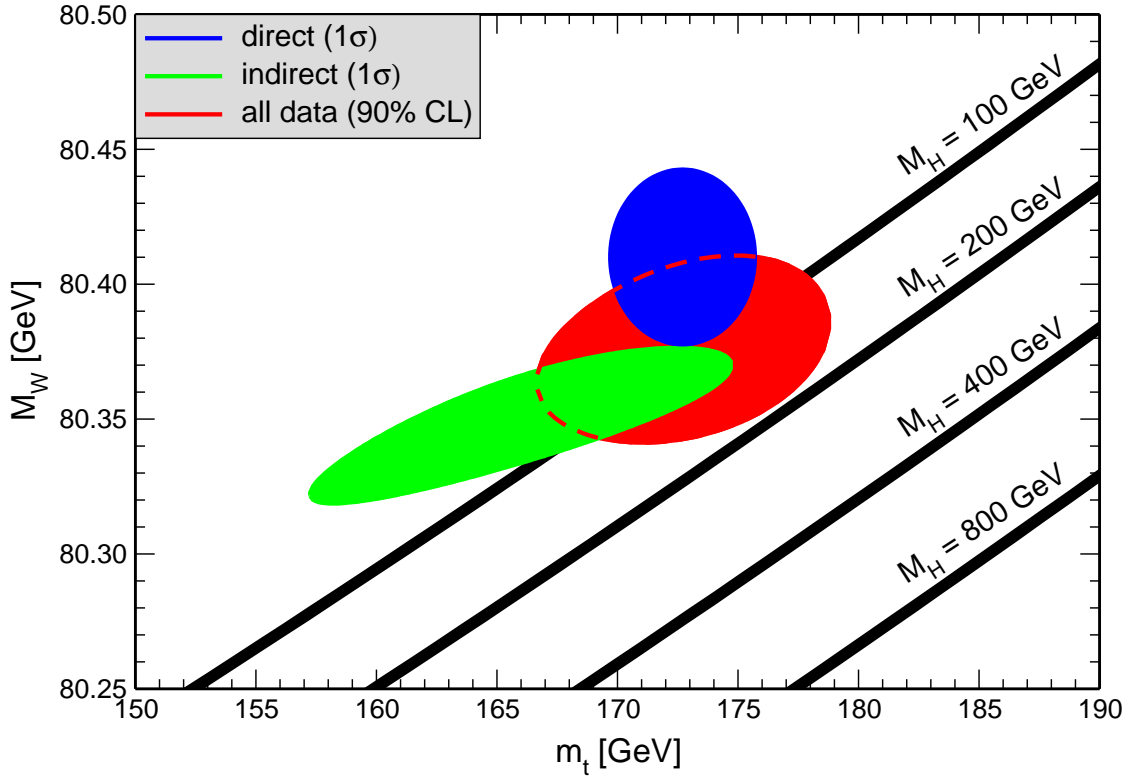}
  \caption{One-standard-deviation (39.35\%) region in $M_W$ as a function of 
           $m_t$ for the direct and indirect data, and the 90\% CL region 
           ($\Delta \chi^2 = 4.605$) allowed by all data. The SM prediction as 
           a function of $M_H$ is also indicated. The width of the $M_H$ bands
           reflects the theoretical uncertainty in the prediction.}
\label{fig:mwmt}
\end{figure}

\begin{figure}[t]
  \includegraphics[height=.47\textheight]{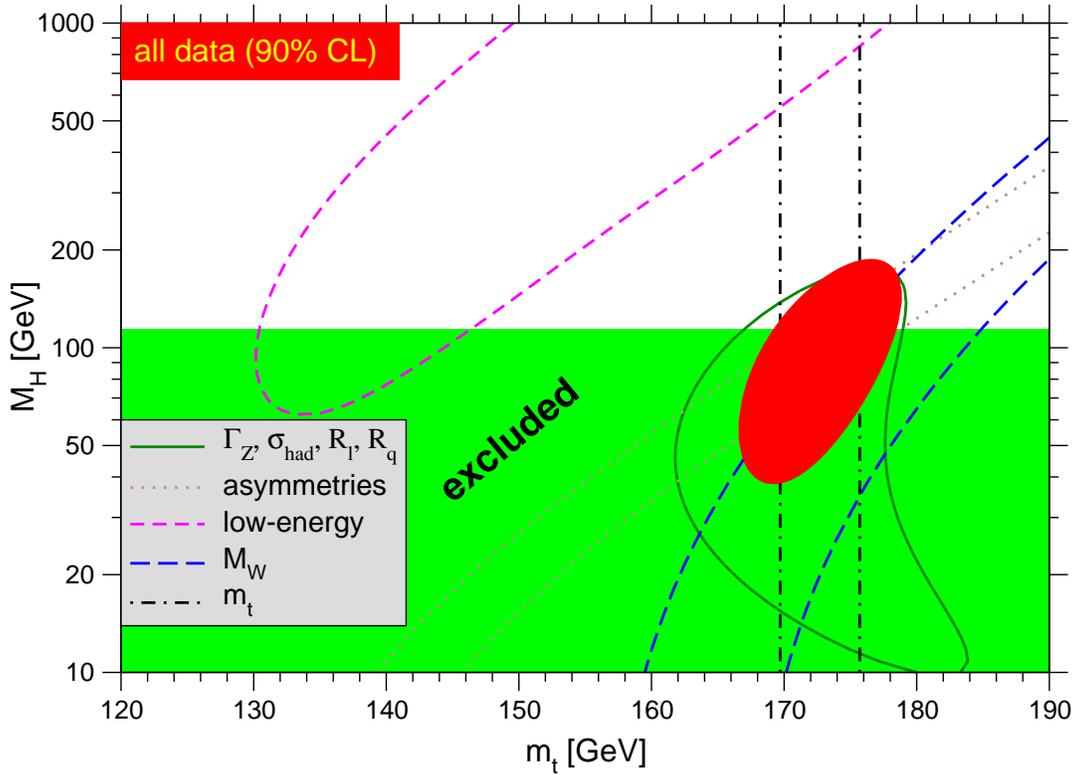}
\caption{One-standard-deviation (39.35\%) uncertainties in $M_H$ as a function
         of $m_t$ for various inputs, and the 90\% CL region allowed by all 
         data. $\alpha_s(M_Z) = 0.120$ is assumed except for the fits including
         the $Z$ lineshape data. The 95\% direct lower limit from LEP~2 is also
         shown.}
\label{fig:mhmt}
\end{figure}

Table~\ref{tab:other} lists non-$Z$ pole observables. Precise values for 
the $W$ boson mass, $M_W$, have been obtained at the high energy frontier at 
LEP~2~\cite{LEP:2005di} ($e^+e^-$) and 
the Tevatron~\cite{Affolder:2000bp,Abazov:2002bu} ($p\bar{p}$). The world 
average, $M_W = 80.410 \pm 0.032$~GeV, has reached a relative precision of 
$4\times 10^{-4}$ with further improvements expected in the near future. 
The direct measurements~\cite{Tevatron:2005cc} of 
$m_t = 172.7 \pm 2.9 \pm 0.6$~GeV from the Tevatron\footnote{The first error 
is experimental~\cite{Tevatron:2005cc} and the second is theoretical from 
the conversion from the top pole mass to the $\overline{\rm MS}$ mass, 
the quantity which actually enters the radiative corrections.} can be compared 
to an indirect determination, $m_t = 172.3^{+10.2}_{-\;\;7.6}$~GeV, from 
the other precision data. The agreement is spectacular. As shown in 
Fig.~\ref{fig:mwmt}, this comparison can even be carried out for the two 
parameters, $m_t$ and $M_W$, simultaneously. In the indirect determination, 
$m_t$ is mostly constrained by $R_b$, but $\Gamma_Z$ and low energy 
measurements also contribute significantly. $M_W$ is then mostly implied by 
the asymmetries. The agreement is again remarkable and it should be stressed 
that there are now two theoretically and experimentally independent indications
for a relatively light Higgs boson with a mass of ${\cal O}(100\mbox{ GeV})$. 
The implications of various sets of observables for $M_H$ and $m_t$ are shown 
in Fig.~\ref{fig:mhmt}.

\begin{table}[t]
\begin{tabular}{llccr}
\hline
  & \tablehead{1}{r}{b}{observable}
  & \tablehead{1}{r}{b}{experimental value}
  & \tablehead{1}{r}{b}{SM prediction}
  & \tablehead{1}{r}{b}{pull}   \\
\hline
&$m_t$          [GeV] & $ 172.7 \pm 3.0   $ & $172.7  \pm 2.8$     & $ 0.0$  \\
&$M_W$          [GeV] & $80.450 \pm 0.058 $ & $80.376 \pm 0.017$   & $ 1.3$  \\
&                     & $80.392 \pm 0.039 $ &                      & $ 0.4$  \\
&$g_L^2$              & $0.30005\pm 0.00137$& $0.30378\pm 0.00021$ & $-2.7$  \\
&$g_R^2$              & $0.03076\pm 0.00110$& $0.03006\pm 0.00003$ & $ 0.6$  \\
&$g_V^{\nu e}$        & $-0.040 \pm 0.015$  & $-0.0396\pm 0.0003$  & $ 0.0$  \\
&$g_A^{\nu e}$        & $-0.507 \pm 0.014$  & $-0.5064\pm 0.0001$  & $ 0.0$  \\
&$A_{PV}$             & $-1.31  \pm 0.17$   & $-1.53  \pm 0.02$    & $ 1.3$  \\
&$Q_W({\rm Cs})$      & $-72.62 \pm 0.46$   & $-73.17 \pm 0.03$    & $ 1.2$  \\
&$Q_W({\rm Tl})$      & $-116.6 \pm 3.7$    & $-116.78\pm 0.05$    & $ 0.1$  \\
&$a_\mu-{\alpha\over 2\pi}$& $4511.07\pm 0.82$ & $4509.82\pm 0.10$ & $ 1.5$  \\
&$\tau_\tau$ [fs]     & $290.89\pm 0.58$    & $291.87\pm 1.76$     & $-0.4$  \\
\hline
\end{tabular}
\caption{Non-$Z$ pole observables compared with the SM best fit predictions. 
The first $M_W$ value is from UA2~\cite{Alitti:1991dk}, 
CDF~\cite{Affolder:2000bp}, and D\O~\cite{Abazov:2002bu}, and the second is 
from LEP~2~\cite{LEP:2005di}. $g_L^2$ and $g_R^2$ (see text) are from 
NuTeV~\cite{Zeller:2001hh}, while the older neutrino deep-inelastic scattering 
($\nu$-DIS) results from CDHS~\cite{Abramowicz:1986vi}, 
CHARM~\cite{Allaby:1986pd}, and CCFR~\cite{Arroyo:1993xx} are included in 
the fits, but not shown. $g^{\nu e}_{V,A}$ are world averaged effective 
four-Fermi couplings in $\nu e$ scattering and dominated by the CHARM~II 
results~\cite{Vilain:1994qy}. $A_{PV}$ is the parity violating asymmetry in 
M\o ller scattering~\cite{Anthony:2005pm}. 
$Q_W({\rm Cs})$~\cite{Wood:1997zq,Guena:2004sq} and 
$Q_W({\rm Tl})$~\cite{Edwards:1995,Vetter:1995vf} are the so-called weak 
charges of Cs and Tl and have been determined in atomic parity violation (APV) 
experiments. The APV errors shown contain significant theory uncertainties 
from atomic structure calculations~\cite{Ginges:2003qt}. In the case of 
$\tau_\tau$ (see text) the theory uncertainty is included in the SM prediction.
In all other SM predictions, the uncertainty is from the SM parameters.}
\label{tab:other}
\end{table}

Other important measurements are from comparatively lower energies or momentum
transfers~\cite{Erler:2004cx}.
The most precise are determinations of anomalous magnetic moments in
leptons, $a_\ell$. The measurement~\cite{Bennett:2004pv} of $a_\mu$ stands out 
because of its unique sensitivity to high energy scales. If the new 
physics~\cite{Czarnecki:2001pv} couples, respectively, through tree or one-loop
effects, a simple dimensional estimate of the scales that can be probed by 
$a_\mu$ at the 1~$\sigma$ level ($\Delta a_\mu$ denotes the total error) gives,
\begin{equation}
   \Lambda_{\rm new} \sim {m_\mu\over\sqrt{\Delta a_\mu}} \sim 3.7 \mbox{ TeV},
   \hspace{50pt} {\Lambda_{\rm new}\over g} \sim 
   {1\over 2\pi} {m_\mu\over\sqrt{\Delta a_\mu}} \sim  590 \mbox{ GeV}.
\end{equation}
The interpretation of $a_\mu$ is complicated by hadronic contributions which 
first arise at the two-loop level. One can use experimental $e^+e^-\rightarrow$
hadrons cross section data to estimate~\cite{Davier:2005xq} the two-loop 
effect, which is due to a vacuum polarization (VP) insertion into a one-loop 
graph, $a_\mu^{(2,{\rm VP})} = (69.54\pm 0.64)\times 10^{-9}$. This value 
suggests a 2.3~$\sigma$ discrepancy between the SM and experiment. If one 
assumes isospin symmetry (which is not exact and appropriate 
corrections~\cite{Cirigliano:2002pv} have to be applied) one can also make use
of $\tau$ decay spectral functions~\cite{Schael:2005am} and one 
obtains~\cite{Davier:2003pw}, 
$a_\mu^{(2,{\rm VP})} = (71.10\pm 0.58)\times 10^{-9}$. This result implies no 
conflict (0.7~$\sigma$) between data and prediction. It is important
to understand the origin of this difference, but the following observations 
point to the conclusion that at least some of it is experimental: 
(i) The latest $e^+e^-$ data by the SND Collaboration~\cite{Achasov:2005rg} are
consistent with the implications of the $\tau$ decay data, and in conflict with
other $e^+ e^-$ data. (ii) The $\tau^- \rightarrow \nu_\tau 2 \pi^-\pi^+\pi^0$ 
spectral function disagrees with the corresponding $e^+ e^-$ data at 
the 4~$\sigma$ level, which translates to a 23\% effect~\cite{Davier:2005xq} 
and seems too large to arise from isospin violation. (iii) Isospin violating 
corrections have been studied in detail in Ref.~\cite{Cirigliano:2002pv} and 
found to be largely under control. The largest effect is due to higher-order 
electroweak corrections~\cite{Marciano:1988vm} but introduces a negligible 
uncertainty~\cite{Erler:2002mv}. (iv) Ref.~\cite{Maltman:2005yk} shows on 
the basis of a QCD sum rule that the spectral functions derived from $\tau$ 
decay data are consistent with values of 
$\alpha_s(M_Z) \buildrel > \over {_\sim} 0.120$ (in agreement with the global 
fit result described in the next section), while the spectral 
functions from $e^+e^-$ annihilation are consistent only for somewhat lower 
(disfavored) values. In any case, due to the suppression at large momentum 
transfer (from where the conflicts originate) these problems are less 
pronounced as far as $a_\mu^{(2,{\rm VP})}$ is concerned, so that it seems
justified to view these differences as fluctuations and to average the results.
An additional uncertainty is induced by the hadronic three-loop light-by-light 
scattering contribution. For this the most recent value,
$a_\mu^{\rm LBLS} = (1.36\pm 0.25) \times 10^{-9}$, of 
Ref.~\cite{Melnikov:2003xd} is employed, which is higher than 
previous evaluations~\cite{Knecht:2001qf,Hayakawa:2001bb}. 

The $\tau$ is the only lepton which can decay hadronically, offering 
a luxurious arena to study the strong interaction and to extract $\alpha_s$. 
Its mass, $m_\tau$, is large enough that the operator product expansion, OPE
(QCD perturbation theory plus almost negligible power corrections in 
an expansion in the inverse $\tau$ mass), can be applied, yet small enough that
QCD effects are sizable 
with great sensitivity to $\alpha_s$. Upon renormalization group evolution from
$m_\tau$ to $M_Z$ (where $\alpha_s$ can be compared to the values from 
$\Gamma_Z$, $\sigma_{\rm had}^0$, and $R_\ell$), the uncertainty scales 
roughly\footnote{This order of magnitude decrease is sometimes called 
the ``incredibly shrinking error''.} like 
$\alpha_s(M_Z)^2/\alpha_s(m_\tau)^2 \sim 0.12$. Furthermore, because the $\tau$
lifetime, $\tau_\tau$, and leptonic branching ratios\footnote{The $\tau$ 
lifetime world average in Table~\ref{tab:other} is computed by combining 
the direct measurements with values derived from the leptonic branching 
ratios.} are fully inclusive, there are no uncertainties from hadronization, 
fragmentation, parton distribution functions, or other modeling of the strong 
interaction. The only potential theoretical uncertainties are from 
the truncation of the perturbative series and from non-perturbative OPE 
breaking effects. The perturbative series is known up to ${\cal O}(\alpha_s^3)$
(the same order as the QCD correction to $\Gamma({\rm had})$) and should 
therefore not be combined with only next-to-leading order determinations of 
$\alpha_s$. The coefficients that enter the $\alpha_s$ expansion of $\tau_\tau$
are relatively large, but dominated by terms that arise from analytical 
continuation and are thus proportional to QCD $\beta$-function coefficients. 
Since the latter are known to ${\cal O}(\alpha_s^4)$ and first enter at 
${\cal O}(\alpha_s^2)$, it is advantageous to treat these effects separately 
and re-sum them to all orders. This amounts to a re-organization of 
the perturbative series (also referred to as ``contour improvement'') with 
smaller expansion coefficients\footnote{These coefficients are given by those 
of the Adler $D$-function, $d_i$.} and where $\alpha_s^n$ is replaced by more 
complicated functions, $A_n(\alpha_s)$. The dominant uncertainty is from 
the lack of knowledge of the four-loop coefficient, $d_3$. One is still exposed
to OPE breaking non-perturbative effects because at one kinematic point one 
needs to change from quark degrees of freedom (QCD) to hadrons (data), but 
fortunately this point is suppressed by a double zero. Very precise data on 
$\tau$ spectral functions (mainly from ALEPH~\cite{Schael:2005am}) constrain 
such effects to a sub-dominant level. 

Currently the largest discrepancy is from $\nu$-DIS scattering. The NuTeV 
Collaboration finds for the on-shell definition of the weak mixing angle, 
$s^2_W = 0.2277\pm 0.0016$, which is 3.0~$\sigma$ higher than the SM 
prediction. The discrepancy is in the left-handed effective four-Fermi 
coupling, $g_L^2 = 0.3000 \pm 0.0014$, which is 2.7~$\sigma$ low, while 
$g_R^2 = 0.0308\pm 0.0011$ is 0.6~$\sigma$ high. At tree level, these are given
by,
\begin{equation}
   g^2_L \approx \frac{1}{2} - \sin^2 \theta_W + \frac{5}{9} \sin^4\theta_W, 
   \hspace{50pt} g^2_R \approx \frac{5}{9} \sin^4 \theta_W.
\end{equation}
Within the SM, one can identify five categories of effects that could cause or 
contribute to this effect~\cite{Davidson:2001ji}: (i) an asymmetric strange 
quark sea, although this possibility is constrained by dimuon 
data~\cite{Goncharov:2001qe}; (ii) isospin symmetry violating parton 
distribution functions at levels much stronger than generally 
expected~\cite{Martin:2003sk}; (iii) nuclear physics 
effects~\cite{Miller:2002xh,Kumano:2002ra}; (iv) QED and electroweak radiative
corrections~\cite{Arbuzov:2004zr,Diener:2005me}; and (v) QCD corrections to 
the structure functions~\cite{Dobrescu:2003ta}. The NuTeV result and the other 
$\nu$-DIS data should therefore be considered as preliminary until 
a re-analysis using PDFs including all experimental and theoretical information
has been completed. It is well conceivable that various effects add up to bring
the NuTeV result in line with the SM prediction. It is likely that the overall 
uncertainties in $g_L^2$ and $g_R^2$ will increase, but at the same time 
the older $\nu$-DIS results may become more precise when analyzed with better 
PDFs than were available at the time.

\section{Global fit}
With these inputs a simultaneous fit to various SM parameters can be performed,
\begin{equation}
\begin{array}{rcl}
M_Z &=& 91.1874 \pm 0.0021~\mbox{GeV}, \\
M_H &=& 89^{+38}_{-28}~\mbox{GeV}, \\
m_t &=& 172.7 \pm 2.8~\mbox{GeV}, \\
\alpha_s (M_Z) &=& 0.1216 \pm 0.0017, \\
\hat\alpha (M_Z)^{-1} &=& 127.904\pm 0.019, \\
\sin^2\hat{\theta}_W &=& 0.23122 \pm 0.00015, \\
s^2_W \equiv 1 - {M_W^2\over M_Z^2} &=& 0.22306 \pm 0.00033,
\end{array}
\end{equation}
where the last two lines show the weak mixing angle in 
the $\overline{\rm MS}$-scheme (coupling based) and the on-shell scheme (vector
meson mass based), respectively. $\hat\alpha (M_Z)$ is the $\overline{\rm MS}$
electromagnetic coupling as it enters at the $Z$ pole. Despite the small
discrepancies discussed in the previous section, the goodness of the fit 
to all data is very good with a $\chi^2/{\rm d.o.f.} = 47.5/42$. 
The probability of a larger $\chi^2$ is 26\%. Experimental correlations have 
been taken into account. Theoretical correlations, {\it e.g.\/} between
$\hat\alpha(M_Z)$ and $g_\mu - 2$ are also addressed\footnote{This is due to 
the common use of the experimental $e^+e^-\rightarrow$ hadrons cross section 
and $\tau$ decay data. The weak mixing angle for momentum transfers at or below
hadronic scales and various hadronic three-loop contributions to $a_\mu$ need
these inputs, as well, implying additional correlations.}. The measurement of 
the latter is higher than the SM prediction, and its inclusion in the fits 
favors a larger $\hat\alpha(M_Z)$ and a lower $M_H$ by about 3~GeV.
 
The extracted $Z$ pole value of $\alpha_s(M_Z)$ is based on a formula with
almost negligible theoretical uncertainty ($\pm 0.0005$ in $\alpha_s (M_Z)$) if
one assumes the exact validity of the SM. One should keep in mind, however, 
that this value\footnote{If one adds non-$Z$ pole observables (other than 
$\tau_\tau$ or $\tau$ leptonic branching ratios), one obtains the slightly
higher value, $\alpha_s = 0.1202 \pm 0.0027$.}, $\alpha_s = 0.1198 \pm 0.0028$,
is very sensitive to such types of new physics as non-universal vertex 
corrections. In contrast, the value derived from $\tau$ decays, 
$\alpha_s(M_Z) = 0.1225^{+0.0025}_{-0.0022}$, is theory dominated but less 
sensitive to new physics. The two values are in remarkable agreement with each 
other. They are also in good agreement with other recent values, such as from 
a 4-jet analysis at OPAL~\cite{Abbiendi:2006vd} ($0.1182 \pm 0.0026$) and from
jet production at HERA~\cite{Specka:2006jr} ($0.1186 \pm 0.0051$), but 
the $\tau$ decay result is somewhat higher than the value, $0.1170 \pm 0.0012$,
from the most recent unquenched lattice calculation~\cite{Mason:2005zx}.
 
\begin{figure}[t]
  \includegraphics[height=.47\textheight]{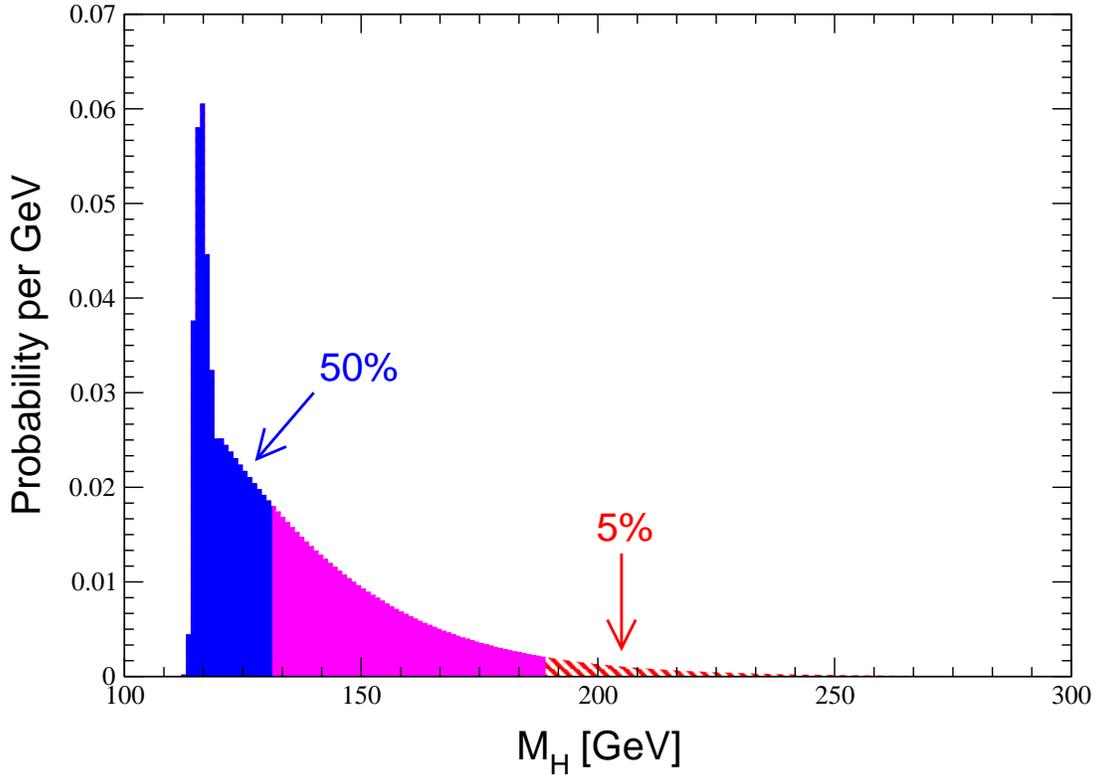}
  \caption{Probability distribution function of $M_H$ including direct search
           results.}
\label{fig:mh}
\end{figure}

There is a strong correlation between the quadratic $m_t$ and logarithmic $M_H$
terms in the radiative corrections except for the $Z\to b\bar{b}$-vertex. $M_W$
has additional $M_H$ dependence which is not coupled to $m_t^2$ effects. 
The strongest individual pulls toward smaller $M_H$ are from $M_W$ and $A_{LR}$
(SLD), while $A^b_{FB}$ and the NuTeV results favor higher values. 
The difference in $\chi^2$ for the global fit is, 
$\Delta\chi^2 = \chi^2(M_H=1000~\hbox{GeV}) - \chi^2_{\rm min} = 60$. Hence, 
the data clearly favor a small value of $M_H$, as in supersymmetric extensions 
of the SM. The 90\% central confidence range from all precision data is
$46 \hbox{ GeV} \leq M_H \leq 154 \hbox{ GeV}$. The central value of the global
fit result, $M_H = 89^{+38}_{-28}$~GeV, is below the direct LEP~2 lower bound,
$M_H \ge 114.4$~GeV (95\%~CL)~\cite{Barate:2003sz}. Including the results of 
these direct searches as an extra contribution to the likelihood function 
drives the 95\% upper limit to $M_H \leq 189$~GeV. As two further refinements, 
the theoretical uncertainties from uncalculated higher order contributions and
the $M_H$-dependence of the correlation matrix which gives slightly more weight
to lower Higgs masses~\cite{Erler:2000cr} are accounted for. The resulting 
limits at 95 (90, 99)\%~CL are
\begin{equation}
   M_H \leq 194 \hbox{ (176, 235) GeV},
\end{equation}
respectively. The probability distribution function of $M_H$ is shown in
Fig.~\ref{fig:mh}.

\section{New Physics and Outlook}
The good agreement between SM predictions and experiments implies strong 
constraints on new physics scenarios beyond the SM.  The $Z$ pole measurements
are particularly suitable to study possible new physics effects on the $Z$
couplings to quarks and leptons.  The per mille precision which has been 
achieved at LEP and SLC allows, for example, only very small mixing between
the $Z$ and a hypothetical extra $Z'$ boson~\cite{Erler:1999ub}. On the other
hand, a $Z'$ with no or little mixing, or other types of new physics 
contributing to $e^+ e^-$ amplitudes without affecting $Z$ boson properties,
could have easily gone unnoticed, since such effects may hide under the $Z$ 
resonance. It is then expedient to examine precision observables away from
the $Z$ pole. High quality and high energy data are provided by
LEP~2~\cite{LEP:2005di}, although these come with comparatively low rates.
An interesting alternative is to utilize low energy observables probing 
directly the weak interaction. This includes processes which exploit the parity
violating character of the weak interaction, as well as neutrino scattering.
 
\begin{figure}[t]
  \includegraphics[height=.47\textheight]{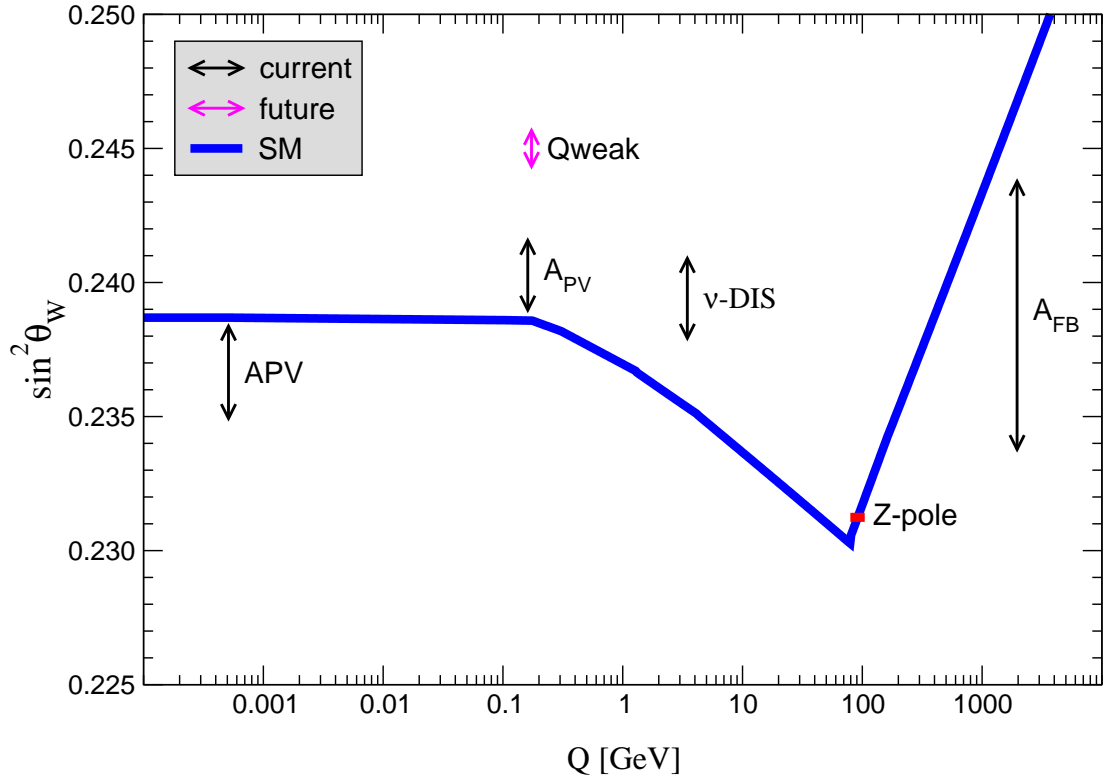}
  \caption{The weak mixing angle in the $\overline{\rm MS}$-scheme as 
           a function of energy, $\sqrt{Q^2}$. The width of the line indicates 
           the uncertainty in the SM prediction. The mixing angle can be 
           determined from a variety of neutral-current processes spanning 
           a very wide $Q^2$ range. The largest discrepancy is the measurement 
           from $\nu$-DIS which is 2.7~$\sigma$ above the prediction. This is 
           mostly due to the NuTeV result~\cite{Zeller:2001hh}. The figure is
           updated from Ref.~\cite{Erler:2004in}.}
\label{fig:s2w}
\end{figure}

For example, the E158 Collaboration at SLAC~\cite{Anthony:2005pm} has extracted
the weak charge of the electron, $Q_W(e)$, from the parity violating asymmetry,
$A_{PV}$, in polarized electron scattering, $\vec{e} e$. A 13\% error in 
the M\o ller asymmetry suffices to access the TeV scale. A similar experiment,
Qweak at JLab~\cite{Armstrong:2003gp}, will determine the analogous proton weak
charge, $Q_W(p)$, in $\vec{e}p$-scattering. The new physics scales probed by 
$Q_W(e)$~\cite{Czarnecki:2000ic} and $Q_W(p)$~\cite{Erler:2003yk} reach (at 
the 1~$\sigma$ level),
\begin{equation}
   {\Lambda_{\rm NEW}\over g} \approx {1\over\sqrt{\sqrt{2} G_F |\Delta Q_W|}}
   \approx \left\{ \begin{array}{c} 4.6\hbox{ TeV } [Q_W(p)], \\ 
                                    3.2\hbox{ TeV } [Q_W(e)],\end{array}\right.
\end{equation}
where $g$ is the coupling strength of the new physics, $G_F$ is the Fermi
constant, and $|\Delta Q_W|$ is the total uncertainty (a 4\% determination of
$Q_W(p)$ is assumed).  The reason for the high reach in these experiments is 
a suppression of the tree-level SM contribution which is proportional to 
$1 - 4\sin^2\theta_W$. The numerical value of $\sin^2\theta_W$ which enters at
very low momentum transfer ($Q^2 \approx 0.03$~GeV$^2$ in these experiments) is
even closer to the ideal value of 1/4 than the one entering $Z$ pole physics
($Q^2 = M_Z^2$). This is illustrated in Fig.~\ref{fig:s2w}.

\begin{figure}[t]
  \includegraphics[height=.47\textheight]{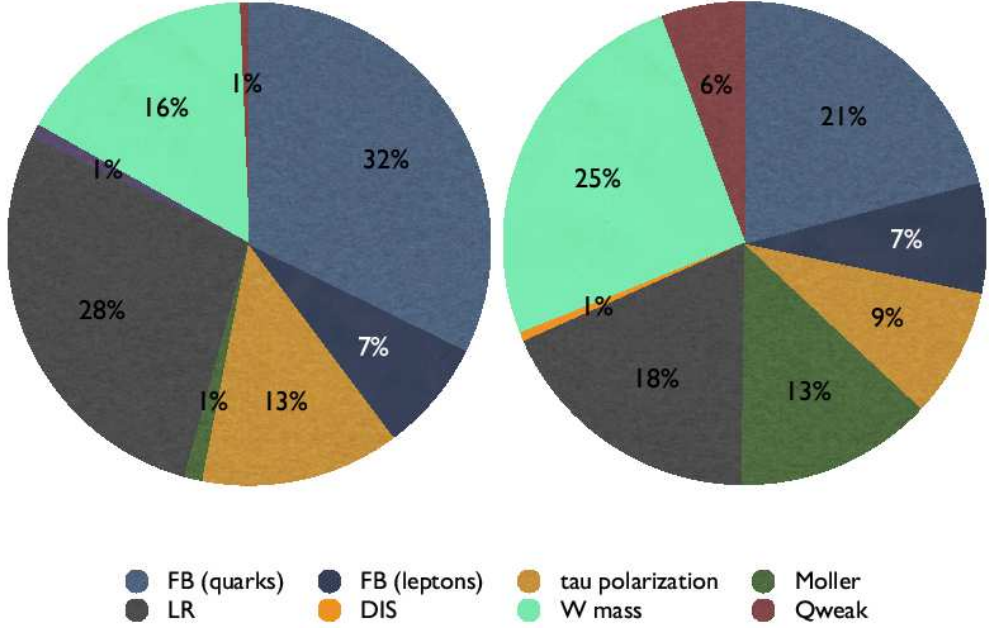}
  \caption{Precision weighted contributors to our knowledge of the weak mixing
           angle, $\sin^2\theta_W$, by type of observable. The left-hand side 
           is the status, while the right-hand side is a projection into 
           the intermediate future, assuming 3.5\% $A_{PV}$ and 4\% $Q_W(p)$
           determinations, as well as 2~fb$^{-1}$ of Tevatron Run~IIA data.
           The $W$ mass can be regarded as a measurement of the on-shell mixing
           angle, $s^2_W$. Qweak refers to the weak
           charges of the proton (Qweak) and heavy nuclei (APV).}
\label{fig:worlds2w}
\end{figure}

\begin{figure}[t]
  \includegraphics[height=.47\textheight]{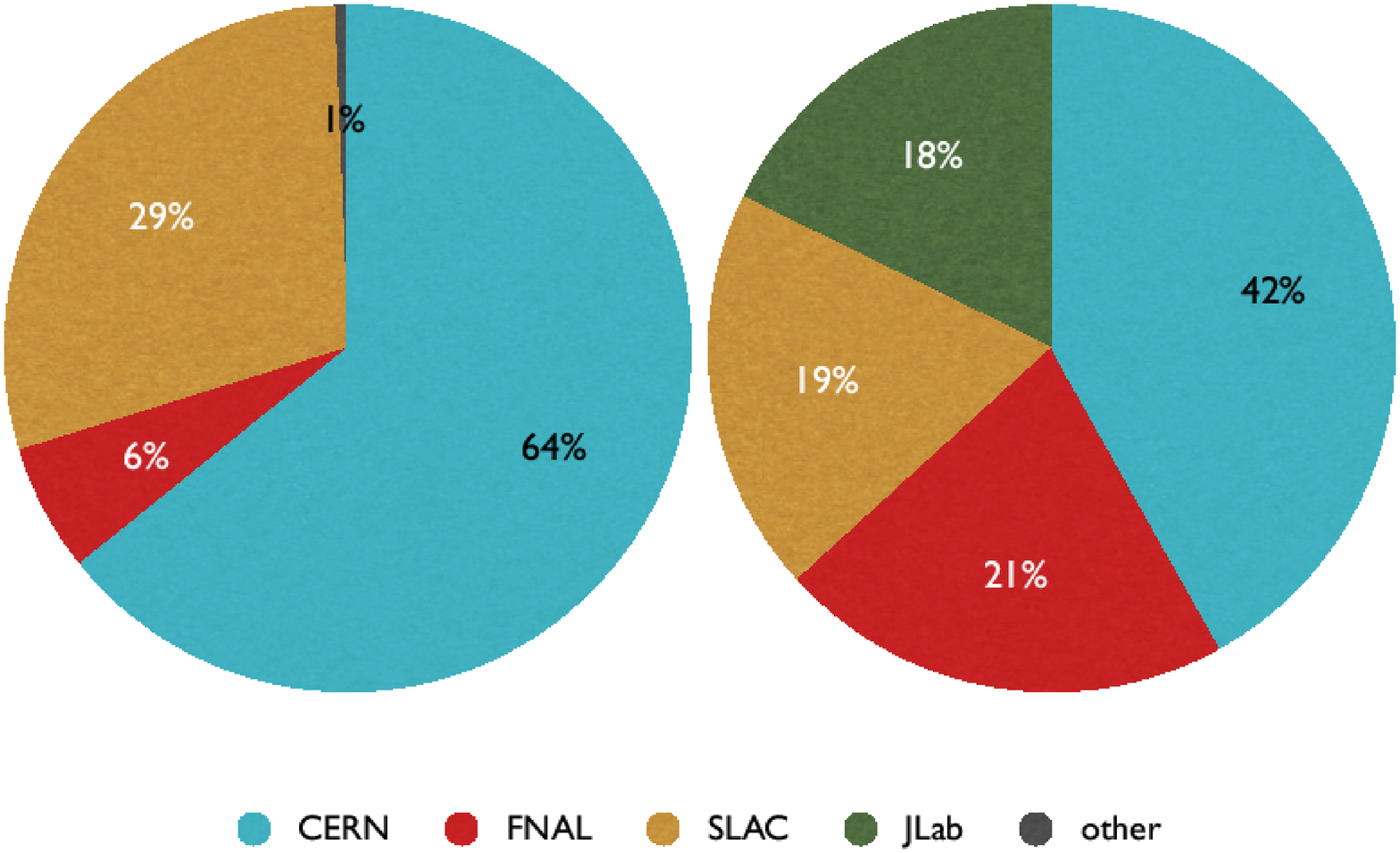}
  \caption{Contributors to our knowledge of the weak mixing angle,
           $\sin^2\theta_W$, by laboratory. The left-hand side is the status, 
           while the right-hand side is the projection using the same 
           assumptions as in Fig~\ref{fig:worlds2w}.}
\label{fig:s2wbyLab}
\end{figure}

This kind of low energy, very high statistics measurement may even compete with
the $Z$ factories. For example, a factor of 4 improvement in $A_{PV}$ relative
to the E158 result (which can conceivably be achieved at an upgraded 12~GeV 
CEBAF at JLab) would yield a measurement of the low energy mixing angle to 
about $\pm 0.00035$. Fig.~\ref{fig:worlds2w} shows a breakdown of our current 
knowledge of the weak mixing angle. A possible projection into the intermediate
future is also shown, where a 3.25\% $A_{PV}$ and a 4\% $Q_W(p)$ measurement 
are assumed, along with some expected improvements~\cite{Baur:2002gp} at 
the Tevatron Run~IIA (corresponding approximately to an accumulated luminosity 
of 2~fb$^{-1}$ of data). It is entertaining to also display what such 
an outcome would mean for the various laboratories. Fig.~\ref{fig:s2wbyLab} 
shows that JLab with a dedicated asymmetry physics program could contribute 
almost as much to $\sin^2\theta_W$ as the high-energy laboratories, SLAC and 
FNAL.

One can also consider the general effects on neutral-current and $Z$ and $W$
boson observables of various types of heavy new physics which contribute to 
the $W$ and $Z$ self-energies but which do not have any (or only small) direct
coupling to the ordinary fermions. In addition to non-degenerate multiplets, 
which break the vector part of $SU(2)_L$, these include heavy degenerate 
multiplets of chiral fermions which break the axial generators. Such effects 
can be described by just three parameters, $S$, $T$, and 
$U$~\cite{Peskin:1991sw}. $T$ is equivalent to 
the electroweak $\rho$-parameter~\cite{Veltman:1977kh} and proportional to 
the difference between the $W$ and $Z$~self-energies at $Q^2=0$ (vector 
$SU(2)_L$-breaking). $S$ and $S+U$ are associated, respectively, with 
the difference between the $Z$ and $W$ self-energies at $Q^2 = M_{Z,W}^2$ and 
$Q^2=0$ (axial $SU(2)_L$-breaking). $S$, $T$, and $U$ are defined with a factor
proportional to $\hat\alpha$ removed, so that they are expected to be of order 
unity in the presence of new physics. A heavy non-degenerate multiplet of 
fermions or scalars contributes positively to $T$, while a multiplet of heavy 
degenerate chiral fermions increases $S$. For example, a heavy degenerate 
ordinary or mirror family would contribute $2/(3\pi)$ to $S$.

\begin{figure}[t]
  \includegraphics[height=.47\textheight]{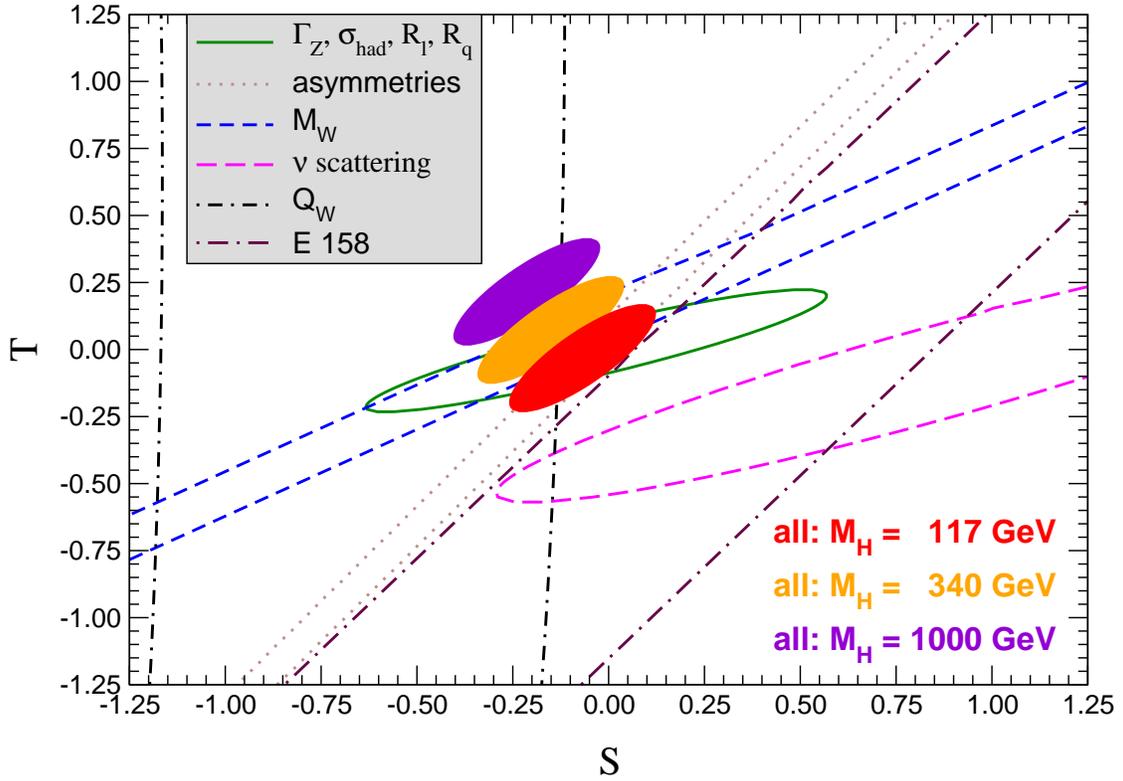}
  \caption{1~$\sigma$ constraints on $S$ and $T$ from various inputs combined 
           with $M_Z$. $S$ and $T$ represent the contributions of new physics 
           only. The contours assume $M_H = 117$~GeV except for the central and
           upper 90\%~CL contours allowed by all data which are for 
           $M_H = 340$~GeV and $1000$~GeV, respectively. $\alpha_s$ is 
           constrained using the $\tau$ lifetime as additional input in all 
           fits.}
\label{fig:ST}
\end{figure}

The data allow a simultaneous determination of $S$, $T$, $U$, and all SM 
parameters except for $M_H$,
\begin{equation}
\begin{array}{rcl}
           S  &=&          -  0.13\pm 0.10\; (-0.08), \\
           T  &=&          -  0.13\pm 0.11\; (+0.09), \\
           U  &=& \phantom{-} 0.20\pm 0.12\; (+0.01), \\
\alpha_s(M_Z) &=&          0.1223 \pm 0.0018,
\end{array}
\label{eq:stu}
\end{equation}
where the uncertainties are from the inputs. The central values assume 
$M_H = 117$~GeV, and in parentheses the change for $M_H = 300$~GeV is shown. 
As can be seen, $\alpha_s$ ($U$) can be determined with no (little) $M_H$ 
dependence. On the other hand, $S$, $T$, and $M_H$ cannot be obtained 
simultaneously, because the Higgs boson loops themselves are resembled
approximately by oblique effects. Eqs.~(\ref{eq:stu}) show that negative 
(positive) contributions to the $S$ ($T$) parameter can weaken or entirely 
remove the strong constraints on $M_H$ from the SM fits. The parameters in 
Eqs.~(\ref{eq:stu}), which by definition are due to new physics only, all 
deviate by more than one standard deviation from the SM values of zero. 
However, these deviations are correlated. Fixing $U = 0$ (as is done in 
Fig.~\ref{fig:ST}) will also move $S$ and $T$ to values compatible with zero 
within errors. Note the strong correlation ($84\%$) between the $S$ and $T$ 
parameters. 

An extra generation of ordinary fermions is excluded at the 99.999\%~CL on 
the basis of the $S$ parameter alone, corresponding to $N_F = 2.81\pm 0.24$ for
the number of families. This result assumes that there are no new contributions
to $T$ or $U$ and therefore that any new families are degenerate. In principle
this restriction can be relaxed by allowing $T$ to vary as well, since $T > 0$
is expected from a non-degenerate extra family.  However, the data currently
favor $T < 0$, thus strengthening the exclusion limits. A more detailed 
analysis is required if the extra neutrino (or the extra down-type quark) is 
close to its direct mass limit~\cite{He:2001tp,Novikov:2002tk}. This can drive
$S$ to small or even negative values but at the expense of too-large 
contributions to $T$. 


\begin{theacknowledgments}
It is a pleasure to thank Paul Langacker and Michael Ramsey-Musolf for fruitful
collaborations. This work was supported by CONACyT (M\'exico) contract 
42026--F and by DGAPA--UNAM contract PAPIIT IN112902.
\end{theacknowledgments}



\bibliographystyle{aipproc}   

\bibliography{sample}

\IfFileExists{\jobname.bbl}{}
 {\typeout{}
  \typeout{******************************************}
  \typeout{** Please run "bibtex \jobname" to optain}
  \typeout{** the bibliography and then re-run LaTeX}
  \typeout{** twice to fix the references!}
  \typeout{******************************************}
  \typeout{}
 }


\end{document}